\documentclass[aps,prb,twocolumn,superscriptaddress,floatfix,amsmath,amssymb]{revtex4-1}

\usepackage{graphicx}
\usepackage{bm}
\usepackage[colorlinks=true,urlcolor=blue,citecolor=blue,linkcolor=blue]{hyperref}
\usepackage{amsmath}

\begin{document}

\title{Critical behaviors of the entanglement and participation entropy
near the many-body localization transition in a disordered quantum spin chain}

\author{Wei Zhang}
\affiliation{Department of Physics, Boston College, Chestnut Hill, MA 02467,
USA}
\author{Ziqiang Wang}
\affiliation{Department of Physics, Boston College, Chestnut Hill, MA 02467,
USA}

\date{\today}

\begin{abstract}
The transition between many-body localized states and the delocalized thermal states is an eigenstate phase transition at finite energy density outside the scope of conventional quantum statistical mechanics. In this work we investigate the properties of the transition by studying the behavior of the entanglement entropy of a subsystem of size $L_A$ in a system of size $L > L_A$ near the critical regime of the many-body localization transition. The many-body eigenstates are obtained by exact diagonalization of
a disordered quantum spin chain under twisted boundary conditions to reduce the finite-size effect.
We present a scaling theory based on the assumption that the transition is continuous and use the subsystem size $L_A/\xi$ as the scaling variable,
where $\xi$ is the correlation length.
We show that this scaling theory provides an effective description of the critical behavior and that the entanglement entropy follows the
thermal volume law at the transition point. We extract the critical exponent governing the divergence of $\xi$ upon approaching the transition point.
We also study the participation entropy in the spin-basis of the domain wall excitations and show that the transition point and the critical exponent agree with those obtained from finite size scaling of the entanglement entropy. Our findings suggest that the many-body localization transition in this model is continuous and describable as a localization transition in the many-body configuration space.

\end{abstract}

\maketitle


\section{\label{sec:level1}Introduction}
In 1958, Anderson proposed that all single particle states of a closed non-interacting quantum system can be localized with sufficient randomness and thus have zero conductivity. Such systems fail to reach thermal equilibrium even after an infinitely long time evolution\cite{Anderson_1958}. About half a century later, Basko, Aleiner, and Altshuler argued that when weak interactions are present, the localization remains\cite{Basko_2006} in the many-body localization (MBL) phase, which has since then been widely studied theoretically and numerically\cite{Pal_2010,Huse_2013, Nandkishore_2014,Nandkishore_2015,Oganesyan_2007,Luitz_2015,Kjall_2014,Bauer_2013,Znidaric_2008,Vosk_2013,Altman_2015,Bardarson_2012,Serbyn_2013_1,Serbyn_2013_2}, and observed experimentally in cold atom and trapped ion systems\cite{Schreiber_2015,Jurcevic_2017,Smith2016,Luschen_2017,Bordia_2016}.

The transition between the MBL phase and the delocalized thermal phase is a dynamical phase transition. As the disorder strength increases, the delocalized system whose long-time behaviour obeys equilibrium thermodynamics turns nonergodic and thus fails to thermalize. Due to the breakdown of ergodicity in the MBL phase, the transition cannot be described by the conventional quantum statistical-mechanics with averages over many eigenstates. In contrast to ordinary quantum phase transitions that occur in ground states, the MBL transition is a transition in the the many-body eigenstates at finite energy densities. The excited eigenstates that satisfy eigenstate thermalization hypothesis (ETH)
are separated from those that fail to satisfy ETH by the MBL transition. A significant implication of ETH is the volume law of the entanglement entropy, whereas it obeys the area law in the MBL phase due to the locality of the interactions, akin to quantum ground states. As a result, the entanglement entropy is widely used as an ``order parameter'' to study the MBL-ETH phase transition.

Although great progresses have been made, some fundamentals of the MBL-ETH phase transition are still not clear, partly because the MBL transition falls outside equilibrium statistical mechanics. Grover argued \cite{Grover_2014} that the critical eigenstates are thermal assuming that the MBL-ETH transition is continuous. While some numerical supports for this analysis have been reported\cite{Kjall_2014, Luitz_2015}, other works suggest the behavior at the critical point to be more like that of a localized phase than an ergodic phase\cite{Pal_2010,Oganesyan_2007}. Arguments against the continuous transition assumption have also proposed\cite{Khemani_2017}. Moreover, following the assumption of a continuous MBL transition, a Harris criterion type of bound $\nu\ge2/d$ for the critical exponent of the divergent length scale
has been proposed\cite{Chandran_2015}. While this bound is corroborated by the perturbative renormalization group (RG) studies\cite{Vosk_2015,Potter_2015,Zhang_2016}, it is violated by essentially {\em all} current exact diagonalization (ED) and scaling of finite-size systems\cite{Kjall_2014,Luitz_2015}. Recently, the validity of the Harris criterion for MBL transition has been challenged and an exact result of $\nu=1$ was derived from the theoretical analysis\cite{Monthus_2016}.
%

Another intriguing and relevant question is wether the MBL ultimately arises through the localization of the many-body states in the configurational Hilbert space of the entire system $L$, in analogy to the single-particle Anderson localization in the physical space. Is the entanglement entropy, which is defined on a subsystem $L_A\in L$, and its volume vs area law in the ETH and MBL phases in the thermodynamic limit $L\gg L_A\gg 1$ just one of the many ways for describing the MBL transition? To be specific, consider an exponentially many expansion coefficients of an eigenstate wave function on the thermal side of the transition over some local basis states. Will the number of expansion coefficients be of lower order on the MBL side? If this is the case, there must exist a quantity defined in that local basis space, in analogy to the inverse participation ratio (IPR) that measures the (inverse) of the spatial coverage of the single-particle eigenstates. This quantity can be called as the many-body IPR (mIPR). Does the mIPR exhibit critical behavior near the MBL transition? The mIRP is clearly basis dependent, but is its critical behavior (if any) basis dependent? Is there a naturally specified choice of basis wherein the MBL transition can be described by the mIRP?
Several theoretical studies have shown that the behavior of the mIPR or its logarithm termed as the participation entropy and the entanglement entropy share similarities\cite{Bera_2015,Beugeling_2015}, but if the former is a critical quantity in the MBL transition is still under debate.

In this work, we report the progress made on a disordered transverse field Ising chain defined in Section IIA, which is known to display the MBL-ETH transition\cite{Kjall_2014}, using an improved ED and a new finite-size scaling analysis that provide several useful insights into these fundamental issues associated with the MBL transition.
Specifically, we apply twisted boundary conditions (TBC) that significantly reduce the finite size effect discussed in Section IIB. The ED is carried out on otherwise identical Ising chains where the end spin is rotated by an angle $\theta$ around $x$ axis. The relevant quantities are averaged over different twisted angles, disorder realizations, and a small energy density window. We find that this algorithm greatly reduces the finite size effect.
We then study the behavior of the entanglement entropy near the transition in Section III, based on the same two assumptions made by Grover\cite{Grover_2014}. The first assumption is that the MBL-ETH transition is continuous. This implies, according to Grover's analysis, that the critical entanglement entropy equals to a thermal entropy. The second is that the entanglement entropy $S_E^A$ of the subsystem is a scaling function only of $L_A/\xi$, with no significant dependence on the total system size $L$ when $L\gg L_A, \xi$. This is reasonable because in the thermodynamic limit, the exact size of the whole system $L$ that acts as a heat bath of the subsystem should not significantly impact the value of $S_A$. We show that $L$ influences $S_E^A$ only through the dimensionless partition ratio $r=L_A/L$. In the relevant thermodynamic limit where $r\rightarrow 0$, we find that $S_E^A$ is strictly thermal at critical point, instead of being subthermal as suggested in Ref\cite{Khemani_2017,Kjall_2014,Luitz_2015}. This is consistent with Grover's analysis and therefore corroborates the assumption of the continuity of $S_E^A$. Following the above analysis, we perform a finite size scaling analysis of $S_E^A$ with $L_A/\xi$ as the scaling variable, whereas $L$ enters through $r$ as corrections to scaling due to irrelevant operators.
In this way we find a critical exponent $\nu=0.94\pm0.07$. This value
still violates the Harris bound, but agrees well with the result derived in Ref\cite{Monthus_2016}. Finally, in Section IV we perform a finite size scaling analysis of the participation entropy, i.e. the logarithm of the mIPR, defined in a suitable spin configuration space of the domain wall excitations. We find that both the critical point and the critical exponent agree with those obtained from the scaling of entanglement entropy. This result implies that MBL-ETH transition is a localized-delocalized transition in the spin configuration space.

\section{\label{sec:level1}Model and Methods}

\subsection{Transverse-field disordered quantum Ising chain}
The quantum transverse-field Ising chain is known to develop the MBL phase when the disorder strength is strong enough. The Hamiltonian of the system is given by~\cite{Kjall_2014}
\begin{equation}
\hat{H}=-\sum_{i=1}^{L-1}J_i\sigma_i^z\sigma_{i+1}^z+
J_2\sum_{i=1}^{L-2}\sigma_i^z\sigma_{i+2}^z+h\sum_{i=1}^L\sigma_i^x
\label{equation:Hamiltonian}
\end{equation}
where $\sigma^{x}$ and $\sigma^{z}$ are Pauli matrices and $L$ is the number of sites in the chain. In Eq.~(\ref{equation:Hamiltonian}), the second nearest neighbor coupling $J_2$ and the transverse external field $h$ are uniform, whereas the nearest neighbor coupling is site-dependent. We use $J_i=J+\delta J_i$, where $J$ is a constant and $\delta J_i$ is randomly taken from a uniform distribution $[-\delta J, \delta J]$. Thus $\delta J$ measures the
disorder strength. For a certain disorder realization, the energy $E$ of the many-body eigenstates of $H$ is bounded within a bandwidth $W= E_{max}-E_{min}$. Consider a disordered ensemble of $H$, the appropriate dimensionless energy is defined by the energy density $\epsilon=2(E-E_{min})/W$ relative to the bandwidth, within a small window around $\epsilon$. All quantities computed later are averaged over different disorder configurations in a small energy window around a fixed $\epsilon=59/60$. We set $J_2=0.5h=0.3 J$, where the ground state of the Hamiltonian has ferromagnetic order in $z$ direction due to $Z_2$ symmetry breaking. As we focus on excited states, in the absence of interactions and randomness, the ferromagnetic order is destroyed at any finite energy density as the excited domain walls are extensive over the whole chain. When there's randomness but no interaction, the domain walls become localized for infinitesimal disorder. With both interaction and randomness, there exists a finite critical disorder strength separating the ETH phase where the domain walls are extensive from the MBL phase where they're localized\cite{Kjall_2014,Nandkishore_2015}.

\subsection{Twisted bondary conditions}
When applying exact diagonalization (ED) to the Hamiltonian to compute the eigen wavefunctions, we normally use open boundary conditions (OBC), namely, ignore the interactions between boundary spins; or periodic boundary conditions (PBC) which allow the boundary spins to couple in the same way as inner spins; or a mixture of the above. If the system size is large enough, such boundary effect will have negligible impact. However, due to the exponential increase of numerical demand, we can only study systems of very limited sizes. $L=16$ is the largest system  size commonly studied using ED for quantum Ising chains. This limitation leads to strong finite size effects, which become more severe for long-ranged interactions.

To reduce the finite size effects, it is common to use twisted boundary conditions (TBC)\cite{Aligia_2000,Thesburg_2014}. In this paper, we implement TBC by rotating the last spin at the end of the Ising chain by an angle $\theta$ around the $x$ axis. This corresponds to a transformation in the Hamiltonian
\begin{equation}
\sigma_L^z \rightarrow e^{i\theta S_L^x/\hbar}\sigma_L^ze^{-i\theta S_L^x/\hbar}
\label{equation:transformation}
\end{equation}
or equivalently,
\begin{eqnarray}
\sigma_1^z\sigma_L^z \rightarrow cos(\theta/2)\sigma_1^z\sigma_L^z+sin(\theta/2)\sigma_1^z\sigma_L^y,\nonumber \\
\sigma_2^z\sigma_L^z \rightarrow cos(\theta/2)\sigma_2^z\sigma_L^z+sin(\theta/2)\sigma_2^z\sigma_L^y,
\label{equation:transformation2}
\end{eqnarray}
respectively in the nearest and second nearest neighbor coupling terms. The coupling term between $\sigma_L^x$ and the transverse field are kept unchanged. When $\theta=0$, the last spin is along the $z$ axis, which is simply PBC; and when $\theta=2\pi$, it correspond to anti-PBC.
We vary $\theta$ from $0$ to $2\pi$. For each $\theta$ value, we use a number of disorder configurations to generate eigen-wavefunctions. All quantities computed from the wavefunctions will be averaged over both different $\theta$ and different disorder configurations within a small energy window around an energy density, which will be discussed in detail below.

\begin{figure}
\includegraphics[trim = 0mm 0mm 0mm 0mm,width=1\columnwidth,clip=true]{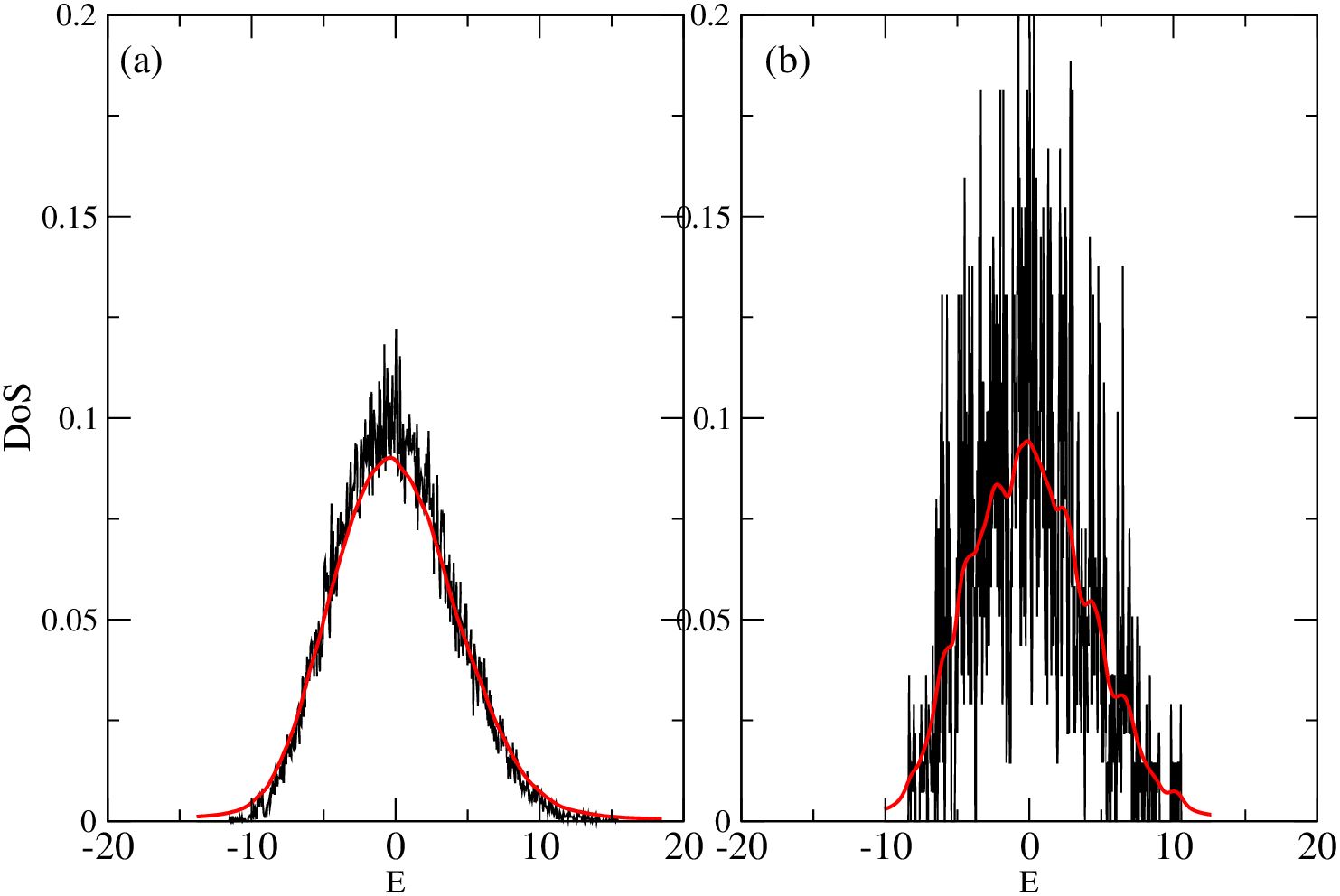}
\caption{Averaged DoS for the disordered spin chain with $L=8$ computed using TBC (a) and PBC (b). Red lines are obtained from spectral function and black lines from counting the number of eigenstates in a small binned energy window. They are obtained by averaging over the same number of ED results and presented under the same parameter settings.}
\label{fig:dos}
\end{figure}

Fig.\ref{fig:dos} shows that the application of TBC greatly improves the smoothness of the density of states (DoS). Fig.\ref{fig:dos}(a) and Fig.\ref{fig:dos}(b) are the DoS obtained using TBC and PBC respectively, with both of them averaged over the same number of ED results. The red line denotes the DoS computed from the spectral function and the black line is obtained by counting the number of eigen-energies in a binned energy window. The values of the infinitesimal real positive number in the spectral function and the width of the energy window are chosen to be the same in Fig.\ref{fig:dos}(a) and Fig.\ref{fig:dos}(b).

\begin{figure}
\includegraphics[trim = 0mm 0mm 0mm 0mm,width=1\columnwidth,clip=true]{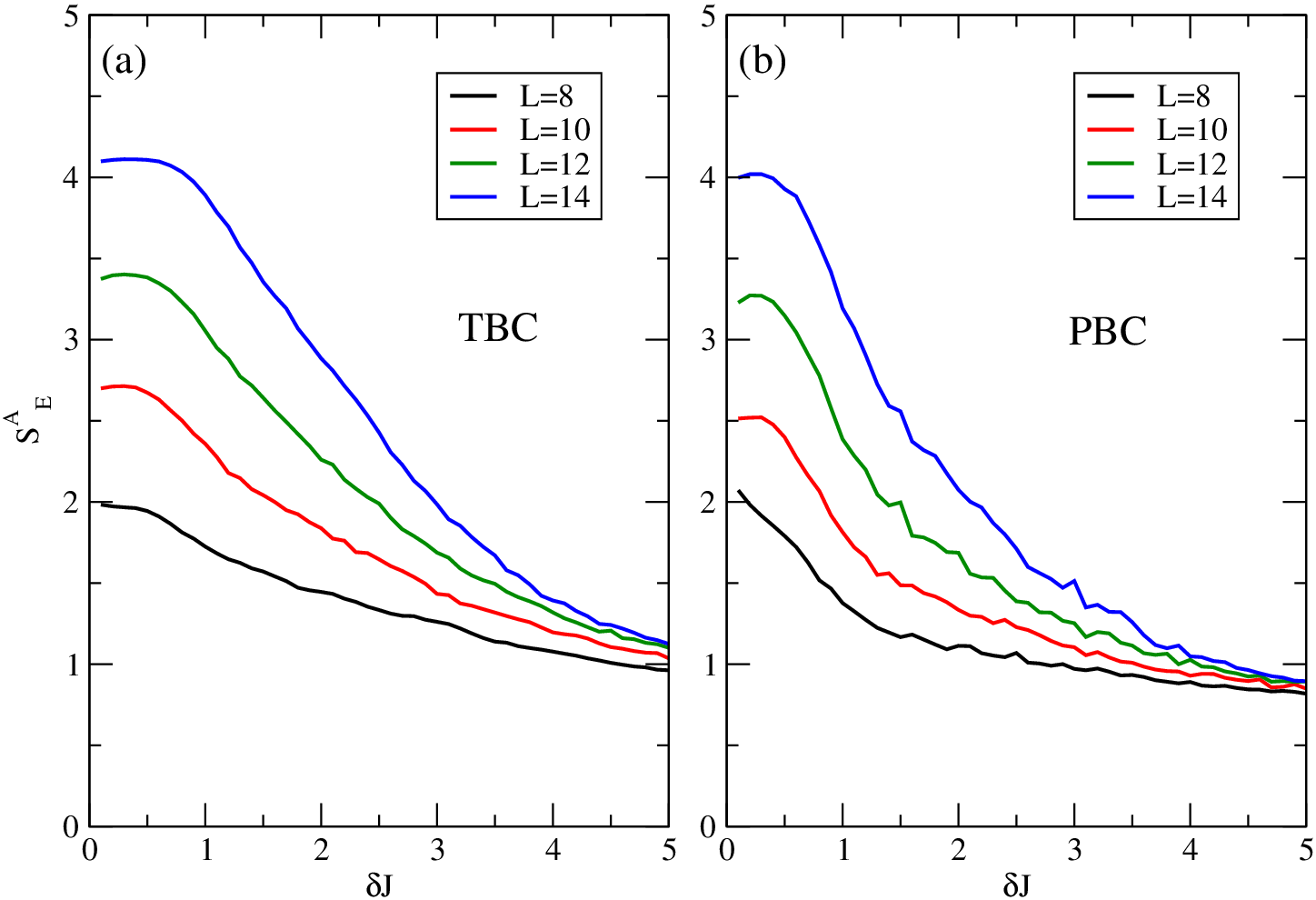}
\caption{Entanglement entropy as a function of disorder strength computed for spin chains of lengths $L=8,10,12,14$ under TBC (a) and PBC (b). The average is taken over the same number of disorder configurations.}
\label{fig:compare_se}
\end{figure}

\begin{figure}
\includegraphics[trim = 0mm 0mm 0mm 0mm,width=1\columnwidth,clip=true]{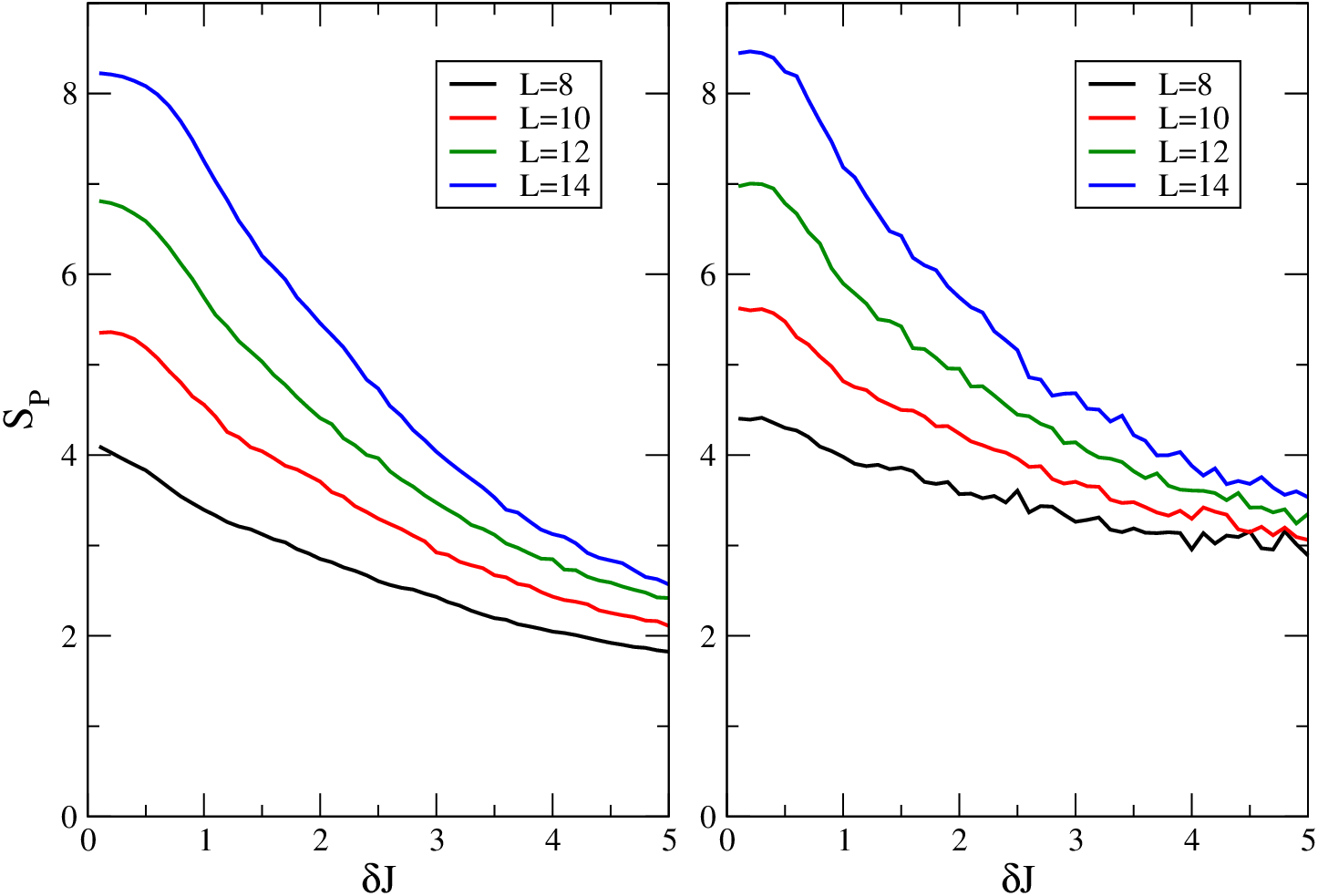}
\caption{Participation entropy as a function of disorder strength computed for spin chains of lengths $L=8,10,12,14$ under TBC (a) and PBC (b). The average is taken over the same number of disorder configurations.}
\label{fig:compare_sr}
\end{figure}

In Fig.\ref{fig:compare_se} and Fig.\ref{fig:compare_sr}, we show that averaging over TBC compared to using pure PBC also reduces the fluctuations of the quantities computed from the wave functions, thus benefits the scaling analysis to be discussed later. We take entanglement entropy and participation entropy (which will be detailed in the following sections) as examples. Fig.\ref{fig:compare_se}(a) shows that the entanglement entropy calculated with TBC has much less fluctuations as a function of the disorder strength, compared to that calculated using PBC in Fig.\ref{fig:compare_se}(b). Fig.\ref{fig:compare_sr} illustrates the same effect for the participation entropy.

From the above results, we find that using TBC can greatly mitigate the boundary effects in finite size systems as it removes the bias of fixing a specific boundary condition and improves the quality of the numerical data by reducing fluctuations of averaged quantities. In the rest of the paper, we will use TBC exclusively to generate all required data.

\section{\label{sec:level2}Entanglement Etropy}

\subsection{Qualitative behavior}

We now study the qualitative behavior of the entanglement entropy $S_E^A(L_A, L, \delta J)$ of a subsystem A of length $L_A$, where $L$ is the length of the whole spin chain and $\delta J$ is the disorder strength. It is given by
\begin{equation}
S_E^A=-\text{Tr}_A\rho_A\text{ln}\rho_A,
\label{sa}
\end{equation}
where 
\begin{equation}
\rho_A=\text{Tr}_{\rm A^c}|\Psi_n\rangle\langle\Psi_n|
\label{rhoa}
\end{equation}
is the reduced density matrix with the trace ${\rm Tr}_{\rm A^c}$ running over the complement set of $A$, i.e. $L-L_A$. $|\Psi_n\rangle$ is the $n^{th}$ eigenstate with energy $E_n$ obtained by exact diagonalization of the Hamiltonian in Eq.~(\ref{equation:Hamiltonian}). In this work we focus on eigenstates with energy densities falling into a small energy window around $\epsilon=59/60$.  It is already clear that in the ETH phase, $S_E^A$ obeys the volume law, namely it is proportional to the {\em subsystem size}. Therefore in our one-dimensional system, it is linear in $L_A$. On the other hand, in the MBL phase, $S_E^A$ should obey the area law, being proportional to the boundary of the subsystem, i.e. it should be a constant in the one-dimensional case. In previous works on the spin chains, it is common to study the half-chain entanglement entropy and its scaling behavior with respect to the whole system size $L$\cite{Kjall_2014,Luitz_2015}. 
Doing so does not clearly addressed if the entanglement entropy scales with $L$ or $L_A$. Although in this case scaling with $L$ is equivalent to scaling with $L_A$, discussing the scaling behavior with respect to the total system size $L$ may be confusing. There are two reasons. First of all, by definition, $S_E^A$ should only scale with $L_A$, since when
$L$ is large enough to satisfy $L\gg L_A$ and $L\gg\xi$ where $\xi\sim |\delta J-\delta J_c|^{-\nu}$ is the correlation length, $S_E^A$ should be independent of the value of $L$. This can be understood since in the thermodynamic limit $L\gg L_A\gg 1$ because if the subsystem in the ETH phase can thermalize and the rest of the system is able to act as an infinite heat bath, the exact size of the heat bath should not matter.
If instead, the subsystem is in the MBL phase, the area law implies that the one-dimensional systems will be independent of either $L_A$ or $L$. Thus the only dimensionless scaling variable should be $L_A/\xi$. Second, requiring $L_A=L/2$ is far away can never really approach the $L_A\ll L$ limit. As a result, the finite size effect may greatly impact the behavior of $S_E^A$ and thus systems may be more difficult to be fully thermalized and thus more prong to subthermal behavior of $S_E^A$. In the ED studied of small to moderate finite size systems, the partition ratio $r=L_A/L$ can enter through corrections to scaling which must be taken into account.
%

\begin{figure}
\includegraphics[trim = 0mm 0mm 0mm 0mm,width=1\columnwidth,clip=true]{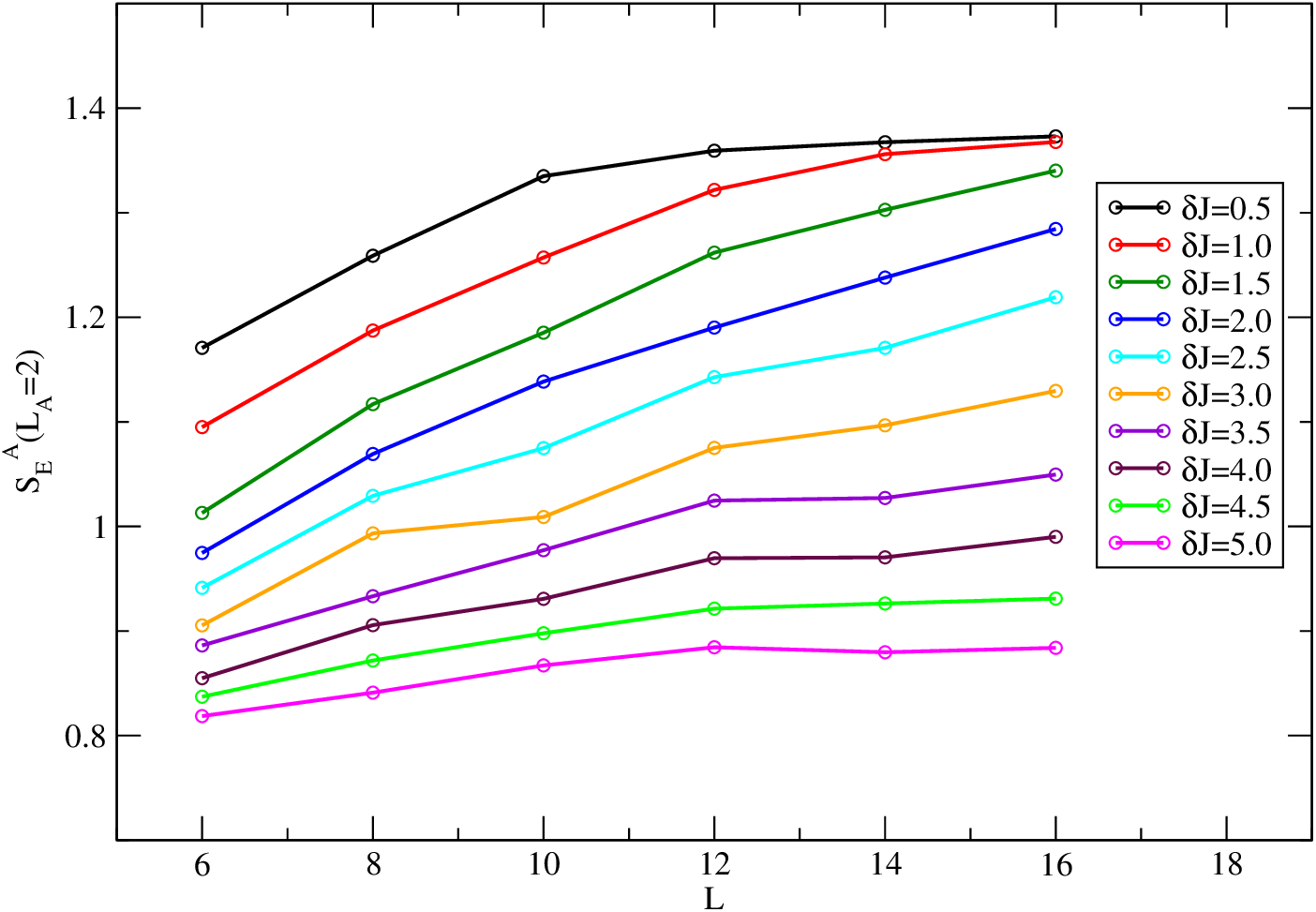}
\caption{The entanglement entropy $S_E^A$ as a function of $L$ at fixed subsystem size $L_A=2$ for different $\delta J$. When $L$ is large, $S_E^A$ is approximately independent of $L$ for both small $\delta J$ smaller than $1.0$) and large $\delta$ (larger than $4.5$). In the region $1.5\le\delta J\le 4.5$, $S_E^A$ shows dependence on $L$ even when for the largest $L=16$.}
\label{fig:se_L}
\end{figure}

\begin{figure}
\includegraphics[trim = 0mm 0mm 0mm 0mm,width=1\columnwidth,clip=true]{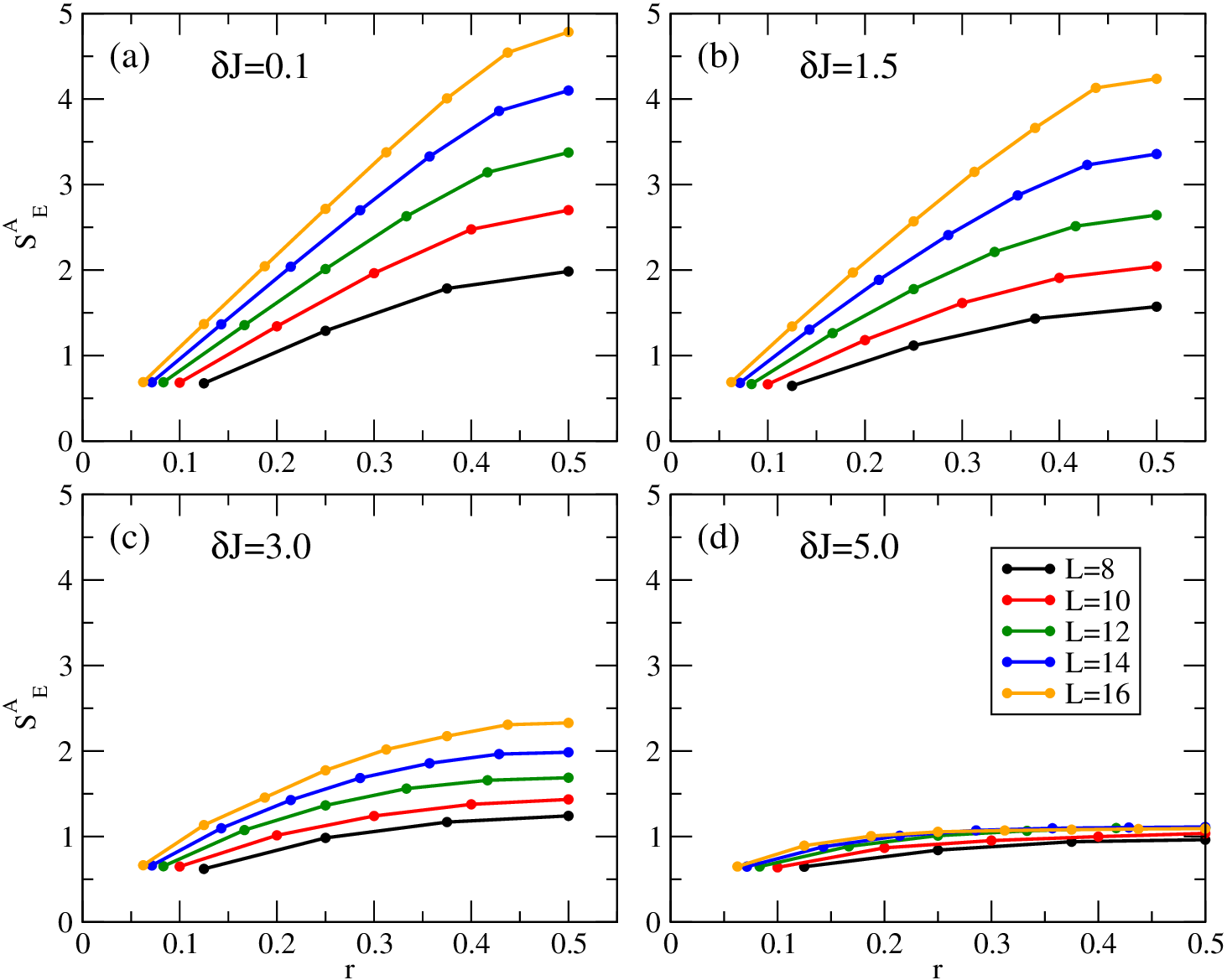}
\caption{Entanglement entropy as a function of the partition ratio $r=L_A/L$ at a fixed $L$ for 5 different system sizes $L=8,10,12,14,16$. Each panel corresponds to a fixed disorder strength: $\delta J=0.1$ (a), $1.5$ (b), $3.0$ (c), and $5.0$ (d). }
\label{fig:se_r_dJ}
\end{figure}

We further illustrate the above statements by calculating the entanglement entropy at fixed $L_A=2$ and explore how $S_E^A$ varies with $L$ for different disorder strength $\delta J$. The results of the ED are shown in Fig.~\ref{fig:se_L}.
It can be seen that when $L>12$, for very weak and very strong disorder strengths, i.e. $\delta J\le 1.0$ and $\delta J\ge 4.5$ respectively, $S_E^A$ shows approximately no dependence on $L$, implying that $L\gg L_A$ and $L\gg\xi$ are both satisfied. On the other hand, if $L\le 12$ in the region of disorder strength above, or for $1.0<\delta J <4.5$ and all $L\le 16$, $S_E^A$ appears to increase with $L$. This latter behavior is likely caused by the violation of $L\gg\xi$, as $\xi$ becomes larger when $\delta J$ gets closer to its critical value. First, we emphasize that this increase is a finite size correction rather than the volume law in $L$. Because $S_E^A$ is bounded in this case by $\text{ln}2L_A$, the dependence of $S_E^A$ on $L$ must vanish for large enough $L$. Second, the finite size correction in the regime where $L$ is comparable to $\xi$ reduces the value of $S_E^A$, which should be accounted for. To achieve this goal, we introduce the partition ratio $r=L_A/L$. When $r\ll 1$, meaning that when $L\gg L_A$, this correction is negligible. When $r$ is finite, it adds a correction to the scaling of $S_E^A$ in the regime around the critical point with $\xi$, in the same spirit as correction to scaling due to irrelevant operators. By doing so we express all $L$ dependence on $S_E^A$ through $r$, thus the entanglement entropy $S_E^A(L_A, L, \delta J)=S_E^A(L_A, r, \delta J)$. The quantitative study of its scaling behavior will be discussed in the next section.
Next, we examine the dependence of $S_E^A$ on $L_A$ and demonstrate why the bipartition entanglement entropy, namely $S_E^A(r=0.5)$, may not be a good choice for studying the scaling behavior of $S_E^A$. In Fig.~\ref{fig:se_r_dJ} we show how $S_E^A$ varies with $r$ for different fixed $L$ at weak, moderate and strong disorder strengths. Fig.~\ref{fig:se_r_dJ}a-c show that at small or moderate $\delta J$, $S_E^A$ shows linear dependence in $r$ for fixed $L$, i.e. linear in $L_A$, when $r$ is small, implying a volume law of $S_E^A$ in $L_A$. As $r$ increases and approaches $0.5$, $S_E^A$'s linear dependence on $L_A$ becomes invalid. The condition that $r$ must be small for the volume law to hold is consistent with the thermodynamic limit requirement $L\gg L_A$, i.e. when the rest of the system can act as an infinite reservoir for the subsystem. Since $S_E^A$ seems not to be fully thermalized at $r=0.5$ due to the small whole system size relative to the subsystem size and do not obey volume law even at very weak disorder strength as seen in Fig.~\ref{fig:se_r_dJ}a, the bipartition entanglement entropy may not be an appropriate choice for studying the scaling behavior. Fig.~\ref{fig:se_r_dJ}d shows that when $\delta J$ is large ($\delta J = 5.0$), the dependence of $S_E^A$ on either $L$ or $L_A$ is rather weak, which is consistent with the area law behavior.

The above observations are qualitatively consistent with the analysis of Grover \cite{Grover_2014} and support the notion that the MBL transition is continuous in the thermodynamic limit.
We note that in ref\cite{Khemani_2017}, it was argued that in the quantum critical regime, the entanglement entropy obeys the area law, which challenges the assumption that the MBL-ETH transition is continuous. We also observed similar plateaus in the $S_E^A$ versus $L$ plot at a small fixed $L_A$ in Fig.~\ref{fig:se_L}. However, since the volume or area law is defined with respect to $L_A$ instead of $L$, we refrain from regarding the plateau as an indication of the area law in quantum critical regime. Moreover, it can be seen in Fig.~\ref{fig:se_r_dJ}c where $\delta J=3.0$, $S_E^A$ for $L=8, 10$ corresponds to the plateau region in Fig.~\ref{fig:se_L}, the entanglement entropy as a function of $L_A$ appears not to behave very differently from those at $L=12, 14, 16$. As a result, we believe the continuity assumption is still valid, and based on which we proceed to build our scaling theory.


\subsection{Finite-size scaling for $S_A$}

The finite-size scaling for the entanglement entropy in the MBL phase transition has been widely studied\cite{Grover_2014,Luitz_2015,Kjall_2014,Khemani_2017,Dumitrescu_2017}. Based on the above observation and discussion, we need to construct a scaling theory explicitly in the subsystem sized $L_A$, such that
when $\delta J<\delta J_c$, $S_E^A$ obeys the volume law in $L_A$ therefore $S_E^A/L_A$ is a finite constant; when $\delta J>\delta J_c$, $S_E^A$ is a constant independent of $L_A$ thus $S_E^A/L_A\rightarrow 0$ when $L_A\gg 1$, following area law in $L_A$; at the critical point $\delta J=\delta J_c$, $S_E^A$ is continuous and thus maintain a critical volume law. To this end, we specify the system and subsystem sizes using $(L_A, r)$ and all correction to the scaling of the entanglement entropy caused by a finite $L$ is expressed in terms of the noncritical dependence on $r$. The entanglement entropy $S_E^A$ thus depends on three variables as $S_E^A(L_A, r, \delta J)$, or equivalently, $S_E^A(L_A, r, \xi(\delta J))$ where $\xi$ is the correlation length. The finite-size scaling form can therefore be written according to
\begin{equation}
S_E^A(L_A,r,\delta J)=L_A f(L_A|\delta J-\delta J_c|^{\nu},r).
\label{equation:scaling_func1}
\end{equation}
The scaling function $f(x,r)$ can be expanded in the vicinity of the critical point,
\begin{equation}
S_E^A/L_A=f_c(r)(1+g_r(L_A|\delta J-\delta J_c|^{\nu})),
\label{equation:scaling_func3}
\end{equation}
where $f_c(r)=f(0,r)$ is the value of $S_E^A/L_A$ at the critical point $\delta J_c$. The function $g_r(x)$ can be expanded as a polynomial in $x$ with $g_r(0)=0$; its expansion coefficients have in general $r$ dependence.

\begin{figure}
\includegraphics[trim = 0mm 0mm 0mm 0mm,width=1\columnwidth,clip=true]{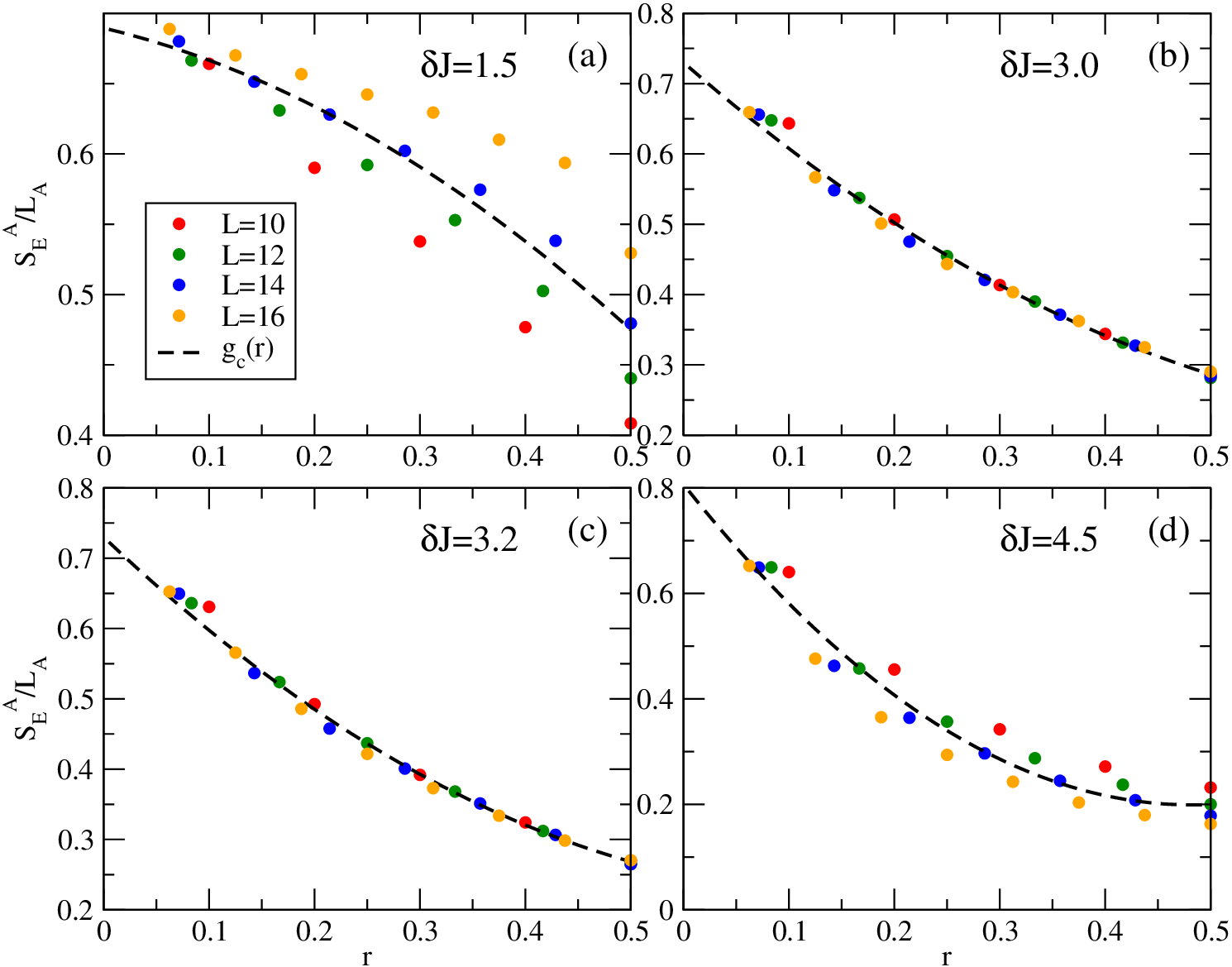}
\caption{Scatter plots of $S_A/L_A$ as a function of $r=L_A/L$ for $\delta J=1.5$ (a), $3.0$ (b), $3.2$ (c), and $4.5$ (d). There are $26$ data points obtained for $L=10,12,14,16$ and $L_A=1,2,\cdots, L/2$. The least squared fits are shown by dashed lines.}
\label{fig:f_c_collapse}
\end{figure}

The functional form in Eq.~(\ref{equation:scaling_func3}) suggests that at the critical point, for all combinations of $L$ and $L_A$,  $S_E^A/L_A$ should collapse to the same function $f_c(r)$. In this way we can determine the critical disorder strength by choosing $\delta J_c$ to be the point where all $S_E^A/L_A$ (for example, all data points in of Fig.\ref{fig:se_r_dJ} divided by their corresponding subsystem size $L_A$) collapse the best. The resulting $f_c(r)$ describes the critical volume law amplitude at the transition for any given $r$. In Fig.~\ref{fig:f_c_collapse}, $S_E^A/L_A$ is shown as a function of $r$ on scatter plots for all $L$ at different disorder strength $\Delta J$. We find that the best data collapse arises at $\delta J_c=3.2\pm0.1$ corresponding to Fig.~\ref{fig:f_c_collapse}c, which gives the critical disorder strength for the transition.

The functional curve of the collapsed data in Fig.~\ref{fig:f_c_collapse}c provides the critical amplitude function $f_c(r)$. Fitting the data according to
\begin{equation}
f_c(r)=\alpha+\beta r+ \gamma r^2
\label{equation:g_c}
\end{equation}
we obtain $\alpha=0.72\pm0.03$, $\beta=-1.31\pm 0.22$, and $\gamma=0.84\pm 0.62$ with the mean square root fitting error equal to $9.98*10^{-3}$. This allows us to extrapolate the critical amplitude to the thermodynamic limit $1\ll L_A \ll L$ by taking the limit,
$\lim_{r\to 0}f_c(r)=a=0.72\pm 0.03$ for the prefactor of the volume-law at critical point. Remarkably, this result agrees within the error bar with the thermal entropy which is $ln2L_A$ in the high temperature limit, where the prefactor is $\text{ln}2\approx 0.693$. This serves as a self-consistency check and confirms that $S_E^A$ is continuous and obeying the volume-law in $L_A$ at transition. Subthermal behavior of $S_E^A$ near the critical point has been observed previously\cite{Kjall_2014,Luitz_2015,Devakul_2015,Lim_2016}, but these works do not focus on the limit $L_A\ll L$. As can be seen from Fig.~\ref{fig:f_c_collapse}c, $f_c(r)<\text{ln}2$ for nonzero $r$, strongly suggesting that the subthermal behavior near the transition is due to a failure to satisfy the $L\gg L_A$ limit that is most severe at the half-chain partition $r=0.5$. In this case, the ability of the rest of the system to act as a heat bath is impaired, making the subsystem not fully thermalized.

We now turn to the finite size scaling analysis of the entanglement entropy $S_E^A$. Instead of scaling $S_E^A/L_A$ in Eq.~(\ref{equation:scaling_func3}), we scale the quantity $y=S_E^A/(L_Af_c(r))$, where $f_c(r)$ has already been obtained by the above procedure and given in Eq.~(\ref{equation:g_c}). Thus $y=1+g_r(x)$ where $x=L_A|\delta J-\delta J_c|^{\nu}$. Since the major dependence of on $r$ has been removed in $y$, we assume the remaining $r$ dependence in the function $g_r(x)$ is small and negligible in the neighborhood of the critical point, provided that $r$ is small and not very close to $0.5$. This assumption can be further discussed below.

\begin{figure}
\includegraphics[trim = 0mm 0mm 0mm 0mm,width=1\columnwidth,clip=true]{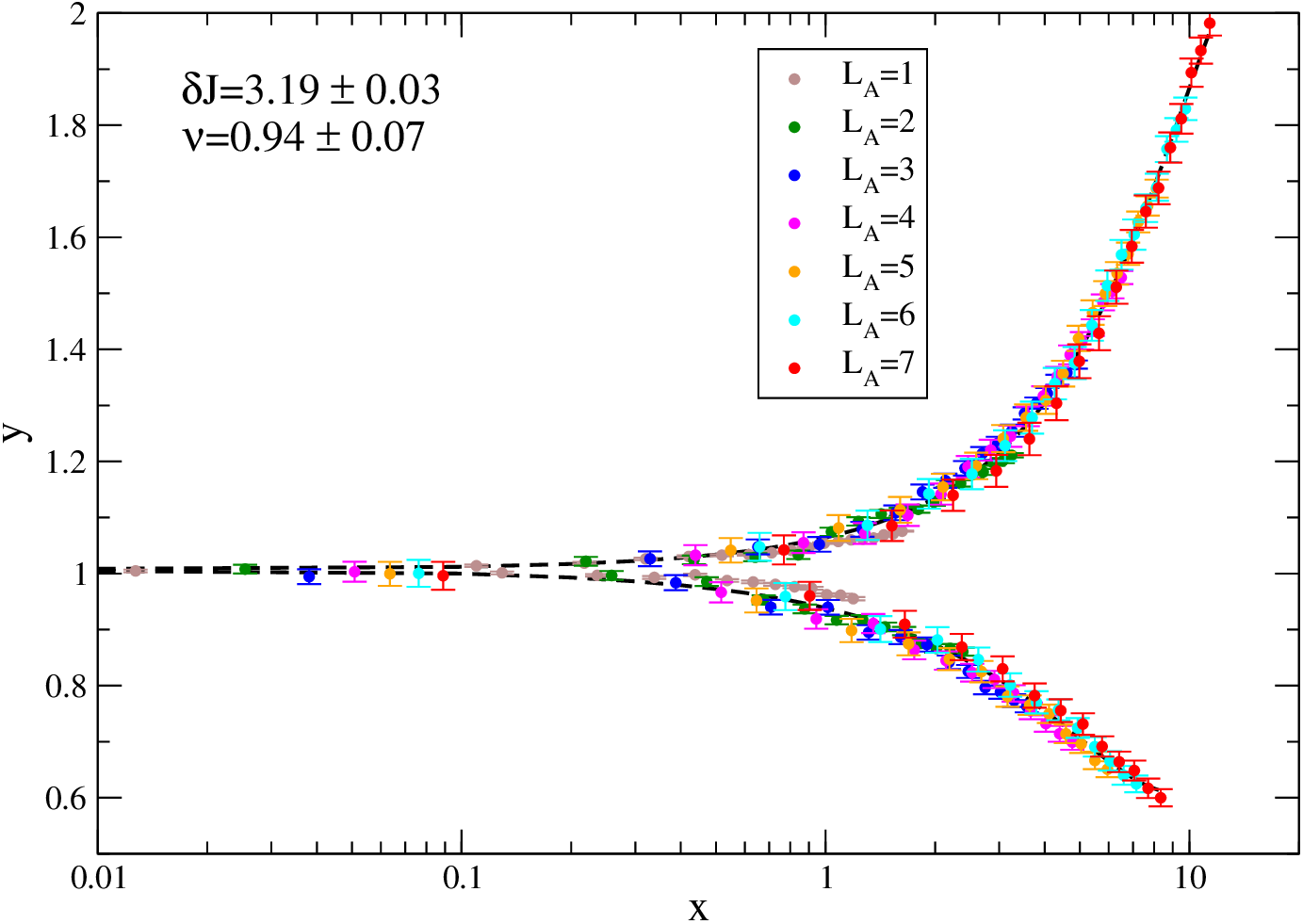}
\caption{Finite size scaling plot of $y=S_E^A/(f_c(r)L_A)$ versus $x$ at $L=16$, where $f_c(r)$ is determined in Fig~\ref{fig:f_c_collapse}c and $x=L_A|\delta J-\delta J_c|^{\nu }$. The scaling function is expanded as a polynomial in $x$.}
\label{fig:fss_se}
\end{figure}

This algorithm allows us to choose the largest system size studied namely $L=16$ and $L_A=1,2,3,4,5,6,7$ to complete the finite size scaling analysis. The obtained scaling plot is shown in Fig.~\ref{fig:fss_se}. The good quality of the data collapse allows us to determine the critical point $\delta J_c=3.19\pm 0.03$, which is consistent with the condition under which $f_c(r)$ was obtained. The critical correlation length exponent is $\nu=0.94\pm 0.067$. Note that
the data for $L_A=8$ are not included, because it corresponds to $r=0.5$ which is too large for the coefficients in $g_r(x)$ to be treated as independent of $r$. We checked that if $L_A=8$ were included in the scaling plot, the quality of the data collapse becomes noticeably poorer. In fact, our theory implied that the value of $L$ should not matter if only we divide $S_E^A/L_A$ by the corresponding $f_c(r)$. To verify this point, we repeated the finite size scaling analysis for $L=14$ and $L_A=1,2,3,4,5,6$ and obtained a similar quality of data collapse with the critical point $\delta J_c=3.14\pm 0.06$ and the correlation length exponent $\nu=0.99\pm 0.09$, which are very close to obtained when using $L=16$. The errors in our analysis are obtained following the approach presented in the supplementary material of reference\cite{Luitz_2015}. The good quality of the data collapse and the similar results obtained for different system sizes justify that the $r$ dependence in $g_r(x)$ near the critical point ($x=0$) is negligible when $r$ is away from $0.5$.


\section{\label{sec:level2}Participation Entropy}

Finally, we investigate if the MBL transition describable by the entanglement entropy of a subsystem $L_A$ in a system $L$ in the limit $1\ll L_A \ll L$ can be characterized by the inverse participation ratio (IPR) or its associated participation entropy in the spin configuration space of the entire system $L$. In other words, is the MBL related to the localization of the eigenstates in the spin configuration space of the entire system $L$ due to strong disorder and correlation?\cite{Torres_2017,Bera_2015,Beugeling_2015,wzhang_2019} The connections between Hilbert space localization and energy level statistics have indeed been explored and demonstrated by experimental measurements recently\cite{Roushan_2017}. The IPR, which is widely used to study the Anderson localization transition\cite{Bell_1972,Wegner_1980}, can be generalized to the many-body probelm\cite{Luitz_2014}, defined for the $n^{th}$ eigenstate $|\Psi_n\rangle$ as,
\begin{equation}
I_{q}(n)=\sum_i|\langle\{\sigma_i^z\}|\Psi_n\rangle|^{2q}, \quad q=2,3,4\cdots,
\label{ipr}
\end{equation}
where $\{\sigma_i^z\}$ is chosen as the basis for the spin configuration for the model parameters used. This choice of basis will be discussed below. The associated participation entropy is given by
\begin{equation}
S_P^q(n)=\frac{1}{1-q}\mathrm{ln}I_q(n).
\label{sp}
\end{equation}
We consider $q=2$ and ignore the superscript $q$ for simplicity in the rest of the discussion.

\begin{figure}
\includegraphics[trim = 0mm 0mm 0mm 0mm,width=1\columnwidth,clip=true]{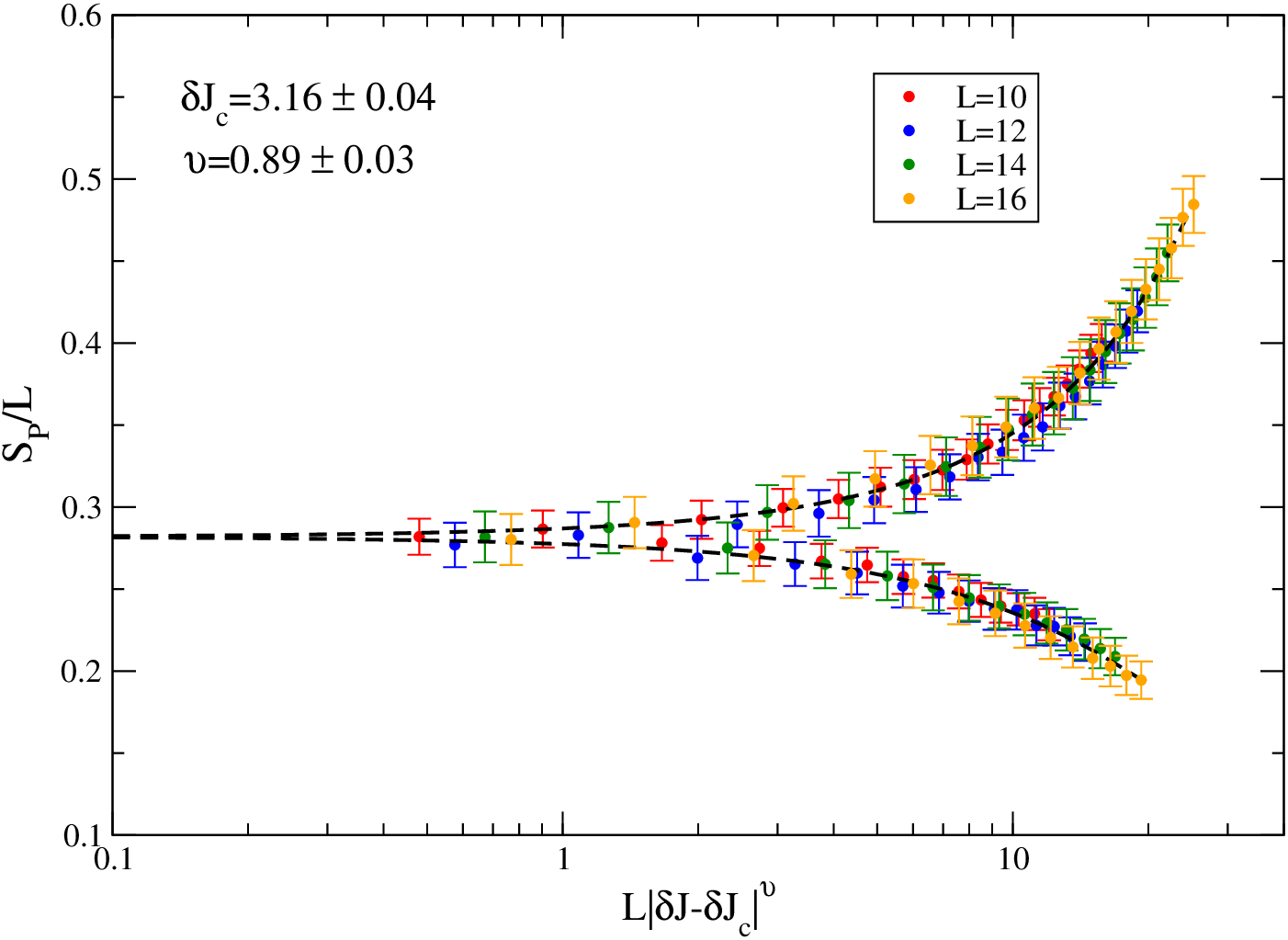}
\caption{Finite size scaling plot of the participation entropy $S_P$ versus $L\vert \delta J-\delta J_c^\prime\vert^{\nu^\prime}$. The critical point determined by the data collapse is $\delta J_c^\prime=3.16\pm0.04$ and the critical exponent $\nu^\prime= 0.89\pm0.03$.}
\label{fig:fss_sr}
\end{figure}

Several theoretical studies have shown that the behavior of $S_P$ and the entanglement entropy share certain similarities\cite{Torres_2017,Bera_2015,Beugeling_2015} and are directly related to each other in the single-particle case\cite{Chen_2012}, while others offer opposite arguments\cite{Luitz_2015,Biroli_2010}. Using the high quality $S_P$ data obtained by ED under the TBC, we performed a finite size scaling analysis of $S_P$ obtained on finite-size systems with $L=10,11,12,13,14,16$ at different disorder strengths. The scaling plot is shown in Fig.\ref{fig:fss_sr}, which gives the critical disorder strength $\delta J_c^\prime=3.16\pm0.04$ and the critical correlation length exponent $\nu^\prime=0.89\pm0.03$. Comparing the results to those obtained for the entanglement entropy$S_A$, we find that $(\delta J_c,\nu)$ and $(\delta J_c^\prime,\nu^\prime)$ agree within the numerical uncertainty, suggesting that the critical behaviors in the partition entropy $S_P$ and the entanglement entropy $S_E^A$ may describe the same phase transition, and therefore MBL is possibly a localization in the spin configuration space.

Note that the definition of the many-body IPR in Eq.~(\ref{ipr}) requires choosing a basis and the obtained results can be basis dependent. It is thus important to discuss if there exists a physically suitable basis for studying MBL using the participation entropy.
Here we chose the spin configuration basis along the $z$ direction in Eq.~(\ref{ipr}) and demonstrated that the participation entropy $S_p$ in Eq.~(\ref{sp}) displays a MBL transition consistent with that of the basis-independent entanglement entropy $S_E^A$ for the spin-chain. The reason for this basis choice is that the ground state of the system has ferromagnetic order of the spins along the $z$ direction at $h=0.6J$. Thus the excited states are domain wall excitations that flip the spin along the $z$ direction\cite{Kjall_2014,Nandkishore_2015}. In the ETH phase the domain wall excitations are extensive, whereas in the MBL phase they are localized. There exists an occupation number representation for the domain walls in terms of the local spin configuration in the $z$ direction. For example, in our model system with $Z_2$ symmetry under $\hat{P}=\prod_{i=1}^L\sigma_i^x$ and OBC, an unoccupied domain wall state $|0\rangle$ is given by $(\vert\uparrow\uparrow\rangle+\vert\downarrow\downarrow\rangle)/\sqrt{2}$, while an occupied domain wall state $|1\rangle$ corresponds to $(\vert\uparrow\downarrow\rangle+\vert\downarrow\uparrow\rangle)/\sqrt{2}$).
We believe that the suitable basis to define the many-body IPR is the quasiparticle occupation number basis. Thus, for the parameter space leading to the ferromagnetic ground state and domain wall excitation, the spin configuration in the $z$ direction is the suitable choice of basis with which we computed the participation entropy. This implies that when $h\gg J$ where the domain wall excitations turn to flip spins in the $x$ direction, the appropriate choice of basis would be the spin configuration in the $x$ direction. While more detailed studies are necessary, we have tested this at $h=4J$ and found that the participation entropy $S_P$ defined in the spin configurations along the $x$ direction shows the scaling behavior, while that along the $z$ direction does not. Finally, any two choices of basis connected by transformations that commute with the Hamiltonian will give the same behavior of the $m$-IPR. In the present case, identical results are obtained if the basis is chosen to be the spin configuration along the $-z$ direction. In the more general case where the Hamiltonian has $SU(2)$ symmetry, the basis choices of spin directions along $x$, $y$ or $z$ will produce the same results.

\section{Summary}

We introduced a finite size scaling theory for the critical behavior of the entanglement entropy near the MBL transition. We emphasized that
it is the subsystem size $L_A$, not the total system size $L$, that cuts off the critical singularity of the MBL transition, such that the only scaling variable is $L_A/\xi$ with $\xi$ being the divergent correlation length. The total system size $L$ enters only as a correction to scaling through the introduction of the partition ratio $r=L_A/L$ that  characterizes the ability of the rest of system to act as an infinite heat bath to the subsystem. While $r$ vanishes in the thermodynamic limit $L\gg L_A$, a large $r$ close to its maximum $r_m=0.5$ pronounces the inability to fully thermalize, leading to subthermal behaviors of the entanglement entropy near the critical point. In the thermodynamic limit, the scaling function produces the volume and area laws of the entanglement entropy in $L_A$ but not in $L$ in the ETH and MBL phases, respectively. We applied this scaling theory to the MBL transition in the disordered transverse field Ising chain. The finite size scaling analysis of the entanglement entropy, obtained using exact diagonalization and twisted boundary conditions to reduce the boundary effects, supports that the MBL-ETH transition is continuous and the entanglement entropy is strictly thermal at the critical point. The correlation length exponent is obtained to be $\nu=0.94\pm0.07$. This value is below the lower bound set by the Harris criteria, as are several other exact diagonalizaiton results. It is, however, close to $\nu=1$ derived by Monthus\cite{Monthus_2016} who argued that the critical exponent for the MBL transition need not satisfy the Harris bound since the quantum many-body state under study experiences $2^L-1$ random energies, which is much more than $L^d$ assumed in the derivation of the Harris bound. It remains to be explored if the Harris bound is violated due to the limited system size, or because it does not apply to the MBL phase transition studied. We also argued that the many-body IPR and the participation entropy defined in the quasiparticle occupation number basis of domain wall excitations in the entire system $L$ can be used to describe the MBL transition as a localization transition of the many-body eigenstates in the spin configuration space. We find that the
finite size scaling analysis of the participation entropy results in a critical point and a correlation length exponent very close to those obtained from the entanglement entropy, suggesting they share similar critical behaviors near the MBL transition. In addition to providing new insights to the subject of MBL, this scaling theory should be applicable to other models of MBL.


\section{acknowledgements}

We thank D.N. Sheng, Xiao Chen, and Lei Wang for many helpful discussions. The work is supported by the U.S. Department of Energy, Basic Energy Sciences Grant No. DE-FG02-99ER45747. The authors thank the Institute of Physics, Chinese Academy of Sciences for hospitality.

\bibliography{main}

\end{document}